\documentstyle[epsf]{article}

\newcommand{\etal}{{et al.}}

\newcommand{\apj}{ApJ}
\newcommand{\apjl}{ApJL}
\newcommand{\apjss}{ApJSS}
\newcommand{\apss}{ApSS}
\newcommand{\mnras}{MNRAS}
\newcommand{\nat}{Nat}
\newcommand{\aap}{A\&A}

\begin{document}

\begin{centering}
{\Large \bf Signatures of helical jets} \\
\vspace*{3mm}

{\large W. Steffen} \\
\vspace*{3mm}

{\it
Department of Physics and Astronomy, University of Manchester,
Schuster Laboratory, Oxford Road, Manchester M13 9PL }

\end{centering}

\begin{abstract}
Observational signatures of helical jets can be found in some X-ray
binaries (XRB), planetary nebulae, Herbig-Haro objects and in jets of
active galactic nuclei (AGN). For the prototypical XRB SS433 a
kinematic model of precessing jets has been applied very successfully
and yielded a determination of its distance which is independent of
conventional methods. In galactic jets precession appears to be the
predominant mechanism for the production of observed helical
signatures. In extragalactic jets other mechanisms seem to be
similarly frequent. As a result of their strong dependence on the
direction of motion with respect to the observer, special relativistic
effects can be pronounced in helical jets. These have to be taken into
account in AGN-jets and the newly discovered galactic sources which
show apparent superluminal motion. Since the galactic superluminal
jets are located in a binary system, jet precession is very likely in
these sources. In this paper I review the main structural and
kinematic signatures of helical jets and briefly mention the physical
mechanisms behind them. I will present kinematic simulations of
relativistic jets which are helically bent or have an internal
helical flow field.
\end{abstract}

\section{Introduction}
\label{intro.sec}

Extragalactic jets show signatures of helical structures on all
observed scales, way down from the sub-parsec up to the kiloparsec
scale \cite{1982ApJ...262..478G,1993ApJ...409..130R,1995A&A...302..335S}.
Similarly, some stellar jets appear to vary their direction of
propagation in regular patterns.  Precessing stellar jets can be
associated with Herbig-Haro objects \cite{1995MNRAS.275..557B,1993MNRAS...260..163R,1992A&A...257..693R,1995A&A...296..431R} or planetary nebulae 
\cite{1993A&A...267..194L}.
However, the prototypical precessing jet is found in the X-ray binary
SS433 \cite{1995yera.conf...12J,1979MNRAS.189P..19M}. The recent
discovery of highly relativistic stellar jets in XRBs
\cite{1995Natur.375..464H,1994Natur.371...46M} with 
indications of wiggling ridge lines has raised hopes that these could
reveal important parameters like distance and precession
period of the binary as it was possible for SS433
\cite{1988ApJ...328..600H}.  Precessing jets have also been invoked as
an explanation for the elusive phenomenon of Gamma Ray Bursters
\cite{1995Ap&SS.231..191F}.

The term `helical jet' is used to describe at least three different
types of jet structures (Fig.\ref{sign.fig}).  First we have {\em
ballistic helical jets} in which the individual fluid elements flow
along straight lines, but with the direction of ejection changing
periodically for different elements such that the instantaneous
overall structure is helical.  The second type is that of {\em
helically bent jets} which are twisted as a whole. In this case all
fluid elements flow along a common twisted path delineated by the
curved jet axis. The third category consists of jets with an {\em
internally helical} structure, which are straight as a whole, but with
the fluid flowing along helical trajectories within the jet. A
further case could be considered in which two or three helically bent
jet branches are braided along a common axis (NGC4258
\cite{1992ApJ...390..365C}, ESO428-G14 \cite{1996ApJL...Falcke}). But
we shall consider these as special cases of helically bent jets.

In this paper I describe the main structural and kinematic signatures
of the three cathegories of helical jets mentioned above. I will
sketch the structural, kinematic, and variability signatures and will
briefly mention some of the underlying physical models and
complications arising from coupling between different processes and
relativistic effects.

\section{General structural and kinematic signatures}
\label{signatures.sec}

One general characteristic of helical structures is that they are not
symmetric with respect to the axis of the helix. The tangential
vectors of the helix on one side are always pointing closer to the
line of sight than on the other. The observed properties like flux,
optical depth, polarization, and others are likely to be different,
especially if the jet plasma is moving at relativistic speeds.  Of
these properties the most straightforward to observe is the
brightness, which is enhanced due to the larger column on the side
with the tangent more aligned with the line of sight. We might find
discrete regularly spaced knots on one side and only weak filaments,
if any, on the other side. In case of a helical internal structure the
column density is the same on both sides, but relativistic motion may
boost the brightness on one side.
 \begin{figure}
 \centering
 \mbox{\epsfxsize=4.8in\epsfbox[0 0 565 700]{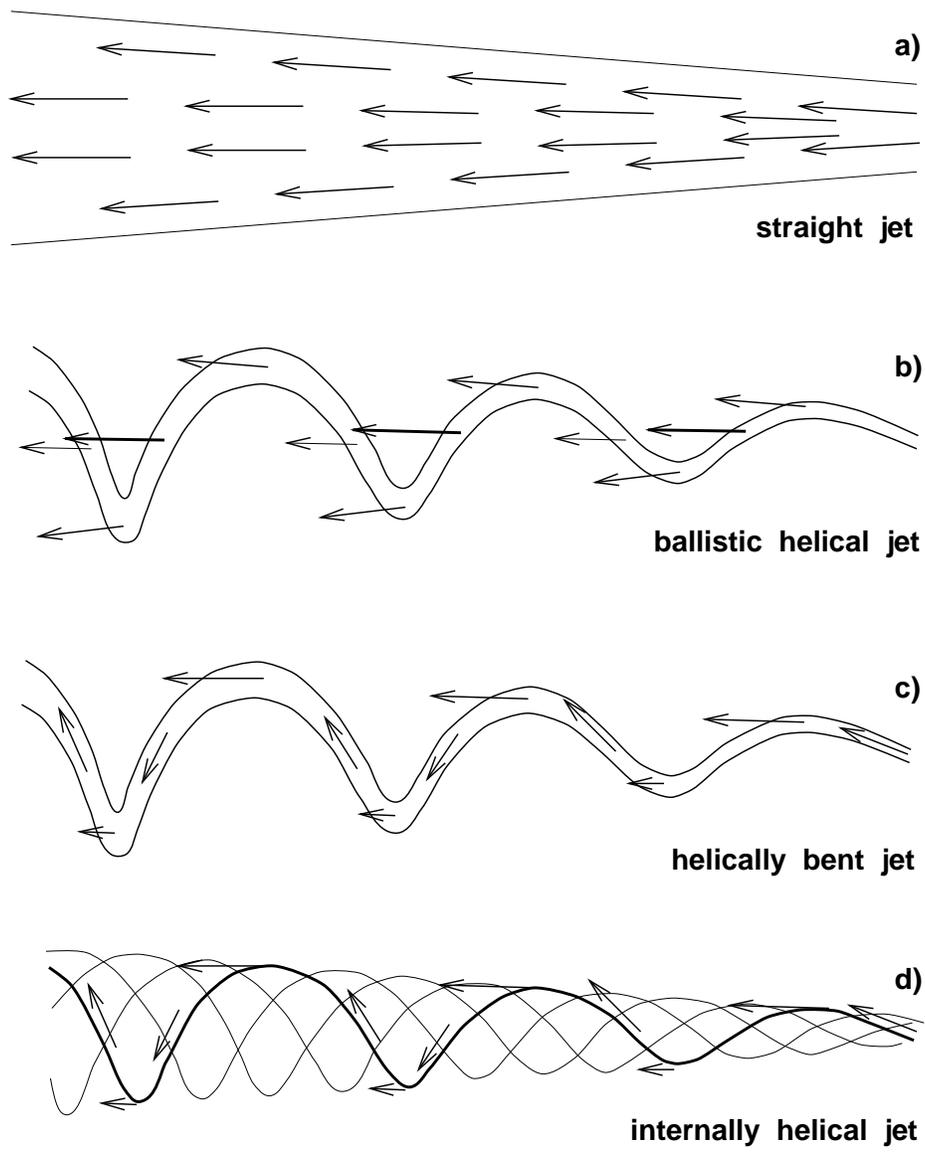}}
 \caption{The proper motion vectors for a straight jet and different kinds of helical jets are shown schematically. }
 \label{sign.fig}
 \end{figure}
\noindent
Three main types of observation can reveal signatures of helical jets:\\

\noindent
a) The projected {\em spatial structure} determined by imaging
   observations; \\ 
b) the {\em motion (kinematics)} deduced from imaging
   or spectroscopy; \\ 
c) the {\em variability} of the object.\\

Useful quantities which can be deduced from these observations are
e.g. the variation of wavelength and amplitude of the oscillations,
knot positions, transverse and radial motions, and brightness
variations, and regular variations of the whole source brightness. For
the discrimination between different possible models the variation of
the regular quantities with distance from the central object may be
useful \cite{1994PhDT...Steffen}.

Relativistic effects can distort the simple geometric features mainly in
two ways. These are light travel-time effects (like apparent superluminal
motion) and differential Doppler-boosting of the emission. Both of these
effect are highly dependent on the direction of motion with respect to the
line of sight. The Doppler-boosting basically increases the brightness contrast
of the knots in a helix, although the effects can be more subtle, like the brightening of one side of a jet with an internal helical flow field (see Section
\ref{helix.sec}).

In Figure \ref{sign.fig} we summarize the observable structural and
kinematic signatures in a schematic diagram.  Figure \ref{sign.fig}a
shows the case of conventional straight jet with the plasma moving on
straight lines roughly parallel to the jet axis.  There are no
external or internal twisted structures nor bents, and the proper
motion of all substructures has roughly the same magnitude and points
to the same direction except for a possibly finite opening angle
$\alpha$.

The single most important parameter determining most of the kinematic
properties of helical jets is the ratio $\eta$ between the densities
of the jet $\rho_j$ and of the external medium $\rho_x$.  ``Heavy''
jet elements with $\eta>1$ propagate almost without resistance through
the interstellar medium and are not deflected from a straight path in
a homogeneous medium \cite{1993MNRAS...260..163R}. Observation of
sinusoidal trajectories of {\em individual} knots is therefore a
strong indication for a helically bent ``light'' jet with $\eta<1$.

\section{Ballistic helical jets}
\label{ballistic.sec}

The case of a ballistic helical jet is illustrated in Figure
\ref{sign.fig}b.  This situation may exist for precessing high Mach number 
jets with densities higher than the environment or which are of lower
density, but highly relativistic.  Here the projected direction of
ejection varies between two limits which define a certain
opening angle of the precession cone $\psi$.  This opening angle has
to be distinguished from an intrinsic opening angle $\alpha$ which the
jet itself may have.

This kind of jet is very much like the moving end of a water hose 
\cite{1993MNRAS...260..163R}:
every drop of water moves independently and radially from the origin
(if no gravity is present) at a velocity near the initial fluid
velocity until it is stopped by the external medium far from the
origin. Because of the opening angle the amplitude of the oscillating
pattern increases with distance, but not the wavelength. The
wavelength at best remains constant, but is more likely to decrease with
distance, because the jet elements will slow down due to the
interaction with the environment.  Therefore, a constant or even
decreasing wavelength combined with radially moving knots is a good
indicator for a precessing jet.

The proper motion of the observed structures will be along straight
lines radially away from the origin of the jet. If the initial
ejection properties of the jet do not vary, the expansion speed at a
given distance from the core will be roughly the same for all knots.
However, because of the finite opening angle the projected proper
motion near the axis of the precession cone may be noticeably
different for the near and the far side. In Fig.\ref{sign.fig} this
has been indicated by proper motion vectors of different lengths.  The
reverse is true for the velocity components along the line of sight,
which may be determined from shifts of spectral lines in optical jets.
Measurement of proper motions, line shifts, and opening
angle of the precession cone allow to determine the true advance speed
and distance to the source \cite{1993A&A...270..177V}.  Cliffe \etal
\cite{1995ApJ...447L..49C} have performed 3D hydrodynamical
simulations of dense strongly precessing jets. Cox, Gull and Scheuer
\cite{1991MNRAS.252..558C} have applied similar simulations with small
precession angles to explain secondary hot spots in radio
galaxies. Extensive kinematic modeling of extragalactic large-scale radio
sources with precessing relativistic jets has been done by Gower
\etal \cite{1982ApJ...262..478G}. The possibly disruptive influence of 
dynamically important magnetic fields on ballistic precessing jets has
be considered by Berry and Kahn \cite{1996MNRAS...BERRY}.

If the precession angle is similar to the jet opening angle, coupling
between the precession and another important process causing helical
jet structures may occur, namely Kelvin-Helmholtz instabilities which
can cause helical twisting of the jet with different characteristics
of growth of amplitude and wavelength
\cite{1987ApJ...318...78H}.

Physical reasons for jet precession may be the wobble of an accretion
disk from which the jet is emerging due to its gravitational
interaction with a secondary mass centre.  The coupling between
unaligned spin axes of the central and the accretion disk can also
result in a precession of the jet. The Lense-Thirring effect, the
interaction between unaligned axis of the orbit and the spin axis of
the body with a jet, can be a reason for precession.  If two jets are
present this will result in a point symmetric structure of the
twin-jets.

Orbital motion of the object with associated jets around a companion
with orbit sizes considerably larger than the jet diameter will cause
the position of the jet ejection in space to change in a circle or
ellipse. The observed ridge line will then be similar to the case with
precession: a wiggling line of emission along a cylinder
without an appreciable opening angle, as opposed to the precessing jet, which
will have a noticable opening angle. The
observed proper motion will be parallel to the axis of this cylinder
and back-extrapolation will generally {\it not} line up with the
source of the jet at its current position. In the case that two jets
are present they will show mirror symmetry with respect to the orbital
plane.

\section{Helical bending}
\label{bent.sec}

In the case discussed above the jet fluid does not flow along the
observed jet structure but rather independently on straight
trajectories into the environment.  However, if the jet as a whole is
helically bent, for whatever reason, and the fluid flows along the
helix, then we encounter different phenomena. If traveling
inhomogeneities are present, both direction and magnitude of their
proper motion vary as they move along the bent trajectory
(Fig.\ref{sign.fig}c). Similarly, the radial velocity changes since the
direction of the velocity vector varies. Note that the variations will
be most pronounced on the side where the local jet axis points closest
to the observer, where traveling and possible stationary components may
merge
\cite{1994A&A...284...51G}.
 \begin{figure}
 \centering
 \mbox{\epsfxsize=4.8in\epsfbox[0 0 421 457]{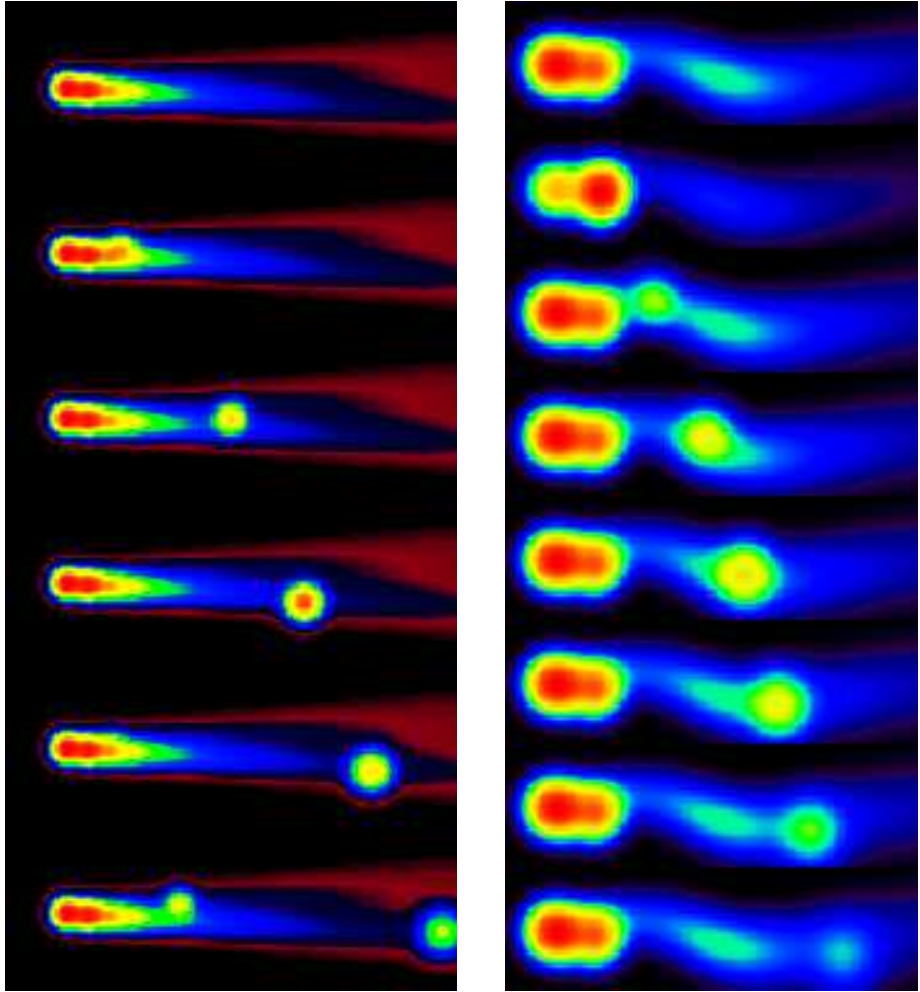}}
 \caption{Time sequences of kinematic simulations of relativistic 
helical jets with
constant opening angles are shown.  The right side is a straight jet
with a helical internal flow field. Note that the emission is
Doppler-boosted on one side of the jet axis. On the right side a
helically bent jet is shown with stationary components at the
positions were the helix points closest to the observer. In both simulations
a plasma inhomogeneity propagates along a helical path through the jet, 
changing brightness due to expansion and differential Doppler effect. 
Both components move with varying apparent superluminal motion.} 
\label{sequence.fig} 
\end{figure}
An important observational signature of helical bending of the jet as
a whole is that all internal structures move along the same path. This
is true as long as the helical pattern can be assumed stationary on
the timescale in consideration. If the propagation of the overall
helical pattern is noticable, the path of consecutive internal
features has to be consistent with the motion of the pattern.  

The motion of the helical pattern should be radially outwards at
roughtly constant speed, similar to what is found in the precession
case, the main difference being that the wavelength and amplitude may
vary in a different manner.  For dynamical reasons, the propagation
speed of the helical pattern as a whole should be considerably smaller
than the internal jet speed. If this is not the case, we are
confronted with a transition case between the helically bent tube and
the ballistic helical jet. This may occur for jets with densities
similar to the external medium, whereas ballistic jets have higher
effective densities and strongly helically bent jets are likely to be
considerably ``lighter'' than the environment, except if magnetic
confinement is very strong.

Jets can be bent by several mechanisms, like e.g. transverse winds,
collisions with clouds, or density gradients. However, none of these
processes provides a helical three-dimensional winding of the jet, but
the bending takes place in a plane. Exceptions can occur in the case
of multiple cloud collisions, which, however, will most likely show a
random zick-zack course with probable jet destruction after a few
interactions \cite{1996ApJSS...233...311H}. Random processes can produce
quasi-regular patterns which can be confused with some characteristics
of helical structures \cite{1986ApJ...300..568R}. The precession
of a ``light'' jet can also produce a helical jet with the plasma
flowing along the curved jet. In this case it has to be taken into
account that the densities which is relevant for the propagation of
the jet upstream from the head of the jet are those of the and the
cocoon cavity which is likely to be formed.

Instabilities of hydrodynamical or magneto-hydrodynamical nature are
able to impose a helical deformation on the surface or the whole body
of the jet. A large number of workers have modelled this behaviour in
two and, more recently, in three dimensions using analytic
calculations or numerical computer simulations 
\cite{1992A&A...256..354A,1989ApJ...342..700C,1986SvAL...12..522G,1987ApJ...318...78H,1992ApJ...400L...9H}.

In Fig.\ref{sequence.fig} (right) a kinematic simulation shows a
plasma inhomogeneity traveling at relativistic speed along a helically
bent jet which itself is also relativistic. It is based on a kinematic
model of the jet in the BL Lac object 1803+78
\cite{1995yera.conf...29S,1995A&A...302..335S,1994PhDT...Steffen}.

\section{Internal helix}
\label{helix.sec}
	
If the fluid inside the jet flows along twisted lines the observed
phenomena may be considerably more complicated compared to the
previous cases. A stationary uniform non-relativistic jet will look
very much like a normal straight jet without internal structure. If
the internal magnetic field in a radio jet is at least to some degree
ordered and follows the helical structure, then using high dynamic
range radio polarization measurements a difference in the 
polarization properties might be found on both sides of the jet axis
\cite{1994A&A...284...51G}. As a result of differential Doppler-boosting 
the side approaching the observer more than the other in a
relativistic jet will appear slightly brighter than the other. The
amount of asymmetry produced by this effect depends strongly on the
Lorentz factor and the twist of the helix. In Fig.
\ref{sequence.fig} (left) a kinematic simulation of this effect on the 
parsec scale of quasar jets is shown. The change in flux density along
the jet axis is the result of adiabatic expansion of the jet
\cite{1994PhDT...Steffen,1995A&A...302..335S}.

The paths of off-axis internal features propagating along different
helical field lines will be different for consecutive features
\cite{1995A&A...302..335S,1995ApJ...443..35Z}. Differential
Doppler-boosting in a relativistic jet will cause the emission to vary
in addition to intrinsic changes (Fig.\ref{sequence.fig}, left)
\cite{1994A&A...284...51G,1996A&A...308..395Q} . Very near the core of
an active galactic nucleus this ``lighthouse effect'' may even cause
quasi-periodic variations of the optical continuum emission
\cite{1993A&A...278..391S}.  Similarly, the proper motions of
individual internal features vary along their path, since each
field line is similar to a helically bent jet as discussed in
Section \ref{bent.sec}.

Most theoretical scenarios for the formation of galactic and
extragalactic jets focus on the symbiosis of a massive and compact
object surrounded by an accretion disk and associated magnetic fields
\cite{1982MNRAS.199..883B}. This seems to be the only viable mechanism
known which can produce highly relativistic and well collimated jets.
Internal helical features described above can be directly associated
with this magnetic field line structure at the base of the jet in this
model for the formation of jets
\cite{1986A&A...156...136C,1995A&A...302..335S}. This mechanism, also
known as the `sling-shot' model, accelerates accretion disk gas along
magnetic field lines which are anchored in the black hole and the
disk.  In this process a helical magnetic field and flow configuration
is set up. In a perfectly cylindrical tube the helical motion of the
jet gas could be maintained for a long time, but only a rather small
opening angle of a few degrees, causes the helical trajectories to
open fairly quickly due to angular momentum conservation
\cite{1992A&A...255...59C,1995A&A...302..335S}.

A different scenario is a force-free configuration of the plasma flow
and the magnetic field as discussed by K\"onigl and Choudhouri
\cite{1985ApJ...289..173K} where a helical mode can dominate the
structure of the jet. K\"onigl and Choudhouri make predictions of
synchrotron emission and its polarization which can be compared with
observations \cite{1990MNRAS.244..750J}.

%\bibliography{helix_ref}
%\bibliographystyle{plain}

%\include{helix.bbl.bak}

\end{document}